\documentclass[aps,twocolumn,amssymb,amsfonts,amsmath,showpacs,final,a4paper,superscriptaddress]{revtex4-2}

\usepackage{float}
\usepackage[dvips,final]{graphicx}
\usepackage{dcolumn}% Align table columns on decimal point
\usepackage{bm}% bold math
\usepackage{textcomp}
\usepackage{amsmath}
\usepackage{color}

\begin{document}

\title{The interplay of local electron correlations and ultrafast spin dynamics in fcc Ni}

\author{Tobias Lojewski}
 \affiliation{Faculty of Physics and Center for Nanointegration Duisburg-Essen (CENIDE), University of Duisburg-Essen, Lotharstr.~1, 47057 Duisburg, Germany}

\author{Mohamed F. Elhanoty}
 \affiliation{Department of Physics and Astronomy, Uppsala University, 75120 Uppsala, Sweden}

\author{Lo\"{i}c Le Guyader}
 \affiliation{European XFEL, Holzkoppel 4, 22869 Schenefeld, Germany}

\author{Oscar Gr\aa n\"as}
 \affiliation{Department of Physics and Astronomy, Uppsala University, 75120 Uppsala, Sweden}

\author{Naman Agarwal}
 \altaffiliation{present address: Department of Physics and Astronomy (IFA), Aarhus University, NY Munkegade 120, 8000 Aarhus C, Denmark}
 \affiliation{European XFEL, Holzkoppel 4, 22869 Schenefeld, Germany}

\author{Christine Boeglin}
 \affiliation{Universit\'{e} de Strasbourg, CNRS, Institut de Physique et Chimie des Mat\'{e}riaux de Strasbourg, UMR 7504, 67000 Strasbourg, France}

\author{Robert Carley}
 \affiliation{European XFEL, Holzkoppel 4, 22869 Schenefeld, Germany}

\author{Andrea Castoldi}
 \affiliation{Dipartimento di Elettronica, Informazione e Bioingegneria, Politecnico di Milano, 20133 Milano, Italy}
 \affiliation{Istituto Nazionale di Fisica Nucleare, Sez., Milano, 20133 Milano, Italy}

\author{Christian David}
 \affiliation{Paul Scherrer Institut, Forschungsstr. 111, 5232 Villigen PSI, Switzerland}

\author{Carsten Deiter}
 \affiliation{European XFEL, Holzkoppel 4, 22869 Schenefeld, Germany}

\author{Florian D\"oring}
 \affiliation{Paul Scherrer Institut, Forschungsstr. 111, 5232 Villigen PSI, Switzerland}

\author{Robin Y. Engel}
 \affiliation{Deutsches Elektronen Synchrotron DESY, 22607 Hamburg, Germany}

\author{Florian Erdinger}
 \altaffiliation{present address: EXTOLL GmbH, 68159 Mannheim, Germany}
 \affiliation{Institute for Computer Engineering, University of Heidelberg, Im Neuenheimer Feld 368, 69120 Heidelberg, Germany}

\author{Hans Fangohr}
 \affiliation{European XFEL, Holzkoppel 4, 22869 Schenefeld, Germany}
 \affiliation{Max-Planck Institute for the Structure and Dynamics of Matter, Luruper Chaussee 149, 22761 Hamburg, Germany}
 \affiliation{University of Southampton, Southampton SO17 1BJ, United Kingdom}

\author{Carlo Fiorini}
 \affiliation{Dipartimento di Elettronica, Informazione e Bioingegneria, Politecnico di Milano, 20133 Milano, Italy}
 \affiliation{Istituto Nazionale di Fisica Nucleare, Sez., Milano, 20133 Milano, Italy}

\author{Peter Fischer}
 \affiliation{Institute for Computer Engineering, University of Heidelberg, Im Neuenheimer Feld 368, 69120 Heidelberg, Germany}

\author{Natalia Gerasimova}
 \affiliation{European XFEL, Holzkoppel 4, 22869 Schenefeld, Germany}

\author{Rafael Gort}
 \affiliation{European XFEL, Holzkoppel 4, 22869 Schenefeld, Germany}

\author{Frank de Groot}
 \affiliation{Materials Chemistry and Catalysis (MCC), Debye Institute for Nanomaterials Science, Utrecht University, Universiteitslaan 99, 3584 CG, Utrecht, The Netherlands}

\author{Karsten Hansen}
 \affiliation{Deutsches Elektronen Synchrotron DESY, 22607 Hamburg, Germany}

\author{Steffen Hauf}
 \affiliation{European XFEL, Holzkoppel 4, 22869 Schenefeld, Germany}

\author{David Hickin}
 \affiliation{European XFEL, Holzkoppel 4, 22869 Schenefeld, Germany}

\author{Manuel Izquierdo}
 \affiliation{European XFEL, Holzkoppel 4, 22869 Schenefeld, Germany}

\author{Benjamin E. Van Kuiken}
 \affiliation{European XFEL, Holzkoppel 4, 22869 Schenefeld, Germany}

\author{Yaroslav Kvashnin}
 \affiliation{Department of Physics and Astronomy, Uppsala University, 75120 Uppsala, Sweden}

\author{Charles-Henri Lambert}
 \affiliation{Department of Materials, ETH Zurich, 8093 Zurich, Switzerland}

\author{David Lomidze}
 \affiliation{European XFEL, Holzkoppel 4, 22869 Schenefeld, Germany}

\author{Stefano Maffessanti}
 \affiliation{Deutsches Elektronen Synchrotron DESY, 22607 Hamburg, Germany}

\author{Laurent Mercadier}
 \affiliation{European XFEL, Holzkoppel 4, 22869 Schenefeld, Germany}

\author{Giuseppe Mercurio}
 \affiliation{European XFEL, Holzkoppel 4, 22869 Schenefeld, Germany}

\author{Piter S. Miedema}
 \affiliation{Deutsches Elektronen Synchrotron DESY, 22607 Hamburg, Germany}

\author{Katharina Ollefs}
 \affiliation{Faculty of Physics and Center for Nanointegration Duisburg-Essen (CENIDE), University of Duisburg-Essen, Lotharstr.~1, 47057 Duisburg, Germany}

\author{Matthias Pace}
 \affiliation{Universit\'{e} de Strasbourg, CNRS, Institut de Physique et Chimie des Mat\'{e}riaux de Strasbourg, UMR 7504, 67000 Strasbourg, France}

\author{Matteo Porro}
 \affiliation{European XFEL, Holzkoppel 4, 22869 Schenefeld, Germany}
 \affiliation{Department of Molecular Sciences and Nanosystems, Ca’ Foscari University of Venice, 30172 Venezia, Italy}

\author{Javad Rezvani}
 \affiliation{Laboratori Nazionali di Frascati, INFN, Via Enrico Fermi 54, 00044 Frascati (Roma), Italy}

\author{Benedikt R\"osner}
 \affiliation{Paul Scherrer Institut, Forschungsstr. 111, 5232 Villigen PSI, Switzerland}

\author{Nico Rothenbach}
 \affiliation{Faculty of Physics and Center for Nanointegration Duisburg-Essen (CENIDE), University of Duisburg-Essen, Lotharstr.~1, 47057 Duisburg, Germany}

\author{Andrey Samartsev}
 \affiliation{European XFEL, Holzkoppel 4, 22869 Schenefeld, Germany}
 \affiliation{Deutsches Elektronen Synchrotron DESY, 22607 Hamburg, Germany}

\author{Andreas Scherz}
 \affiliation{European XFEL, Holzkoppel 4, 22869 Schenefeld, Germany}

\author{Justina Schlappa}
 \affiliation{European XFEL, Holzkoppel 4, 22869 Schenefeld, Germany}

\author{Christian Stamm}
 \affiliation{Department of Materials, ETH Zurich, 8093 Zurich, Switzerland}
 \affiliation{Institute for Electric Power Systems, University of Applied Sciences and Arts Northwestern Switzerland, 5210 Windisch, Switzerland}

\author{Martin Teichmann}
 \affiliation{European XFEL, Holzkoppel 4, 22869 Schenefeld, Germany}

\author{Patrik Thunstrom}
 \affiliation{Department of Physics and Astronomy, Uppsala University, 75120 Uppsala, Sweden}

\author{Monica Turcato}
 \affiliation{European XFEL, Holzkoppel 4, 22869 Schenefeld, Germany}

\author{Alexander Yaroslavtsev}
 \affiliation{Department of Physics and Astronomy, Uppsala University, 75120 Uppsala, Sweden}
 \affiliation{European XFEL, Holzkoppel 4, 22869 Schenefeld, Germany}

\author{Jun Zhu}
 \affiliation{European XFEL, Holzkoppel 4, 22869 Schenefeld, Germany}

\author{Martin Beye}
 \affiliation{Deutsches Elektronen Synchrotron DESY, 22607 Hamburg, Germany}

\author{Heiko Wende}
 \affiliation{Faculty of Physics and Center for Nanointegration Duisburg-Essen (CENIDE), University of Duisburg-Essen, Lotharstr.~1, 47057 Duisburg, Germany}

\author{Uwe Bovensiepen}
 \affiliation{Faculty of Physics and Center for Nanointegration Duisburg-Essen (CENIDE), University of Duisburg-Essen, Lotharstr.~1, 47057 Duisburg, Germany}

\author{Olle Eriksson}
 \affiliation{Department of Physics and Astronomy, Uppsala University, 75120 Uppsala, Sweden}
 \affiliation{School of Science and Technology, \"{O}rebro University, 70182 \"{O}rebro, Sweden}

\author{Andrea Eschenlohr}
 \email[]{andrea.eschenlohr@uni-due.de}
 \affiliation{Faculty of Physics and Center for Nanointegration Duisburg-Essen (CENIDE), University of Duisburg-Essen, Lotharstr.~1, 47057 Duisburg, Germany}

\date{\today}

\begin{abstract}
The complex electronic structure of metallic ferromagnets is determined by a balance between exchange interaction, electron hopping leading to band formation, and local Coulomb repulsion. The interplay between the respective terms of the Hamiltonian is of fundamental interest, since it produces most, if not all, of the exotic phenomena observed in the solid state. By combining high energy and temporal resolution in femtosecond time-resolved X-ray absorption spectroscopy with \textit{ab initio} time-dependent density functional theory we analyze the electronic structure in fcc Ni on the time scale of these interactions in a pump-probe experiment. We distinguish transient broadening and energy shifts in the absorption spectra, which we demonstrate to be caused by electron repopulation and correlation-induced modifications of the electronic structure, respectively. Importantly, the theoretical description of this experimental result hence requires to take the local Coulomb interaction into account, revealing a temporal interplay between band formation, exchange interaction, and Coulomb repulsion.
\end{abstract}

\maketitle

Magnetic order in the $3d$ transition metals Fe, Co, Ni and their alloys arises from the effects of exchange interaction, local correlations, and the electronic band structure (electronic hopping) \cite{suhl2012magnetism}. Solid state spectroscopy in conjunction with advanced electronic structure calculations \cite{sanchez2012effects} resolves the underlying microscopic processes in the thermodynamic ground state \cite{stohr2006magnetism} as well as electron, spin, and lattice excitations \cite{schafer2004electronic, kittel1953relaxation}. For fcc Ni, the imaginary part of the self energy representing the scattering rate $\Gamma$ increases from the Fermi energy $(E_{\mathrm F})$ up to 2~eV above  from 100~meV to 500~meV \cite{sanchez2012effects} or, following $\tau=\hbar/\Gamma$, the scattering time $\tau$ decreases from 6 to 1 fs. Time domain methods probe these ultrafast timescales directly and have revealed the optically induced ultrafast demagnetization of the $3d$ ferromagnets \cite{beaurepaire_ultrafast_1996, kirilyuk2010ultrafast}. The underlying processes are based on spin-orbit mediated spin flips \cite{koopmans_2010,toews_pastor_2015,krieger_2015}, spin transfer \cite{battiato_2012,dewhurst_laser-induced_2018,hofherr_2020}, spin-lattice coupling, and the principle of angular momentum conservation \cite{dornes2019ultrafast, tauchert2022polarized}. However, a comprehensive picture of the transient electronic structure is still lacking because the competition (or cooperation) of magnetic order, local correlations, and the optical excitation in a regime beyond a weak perturbation cannot yet fully be accounted for.

In this Letter, we establish the influence of local electronic Coulomb interactions on the spin-dependent electron dynamics in fcc Ni in the time domain. This finding is based on exploiting high energy resolution in fs time-resolved X-ray absorption spectroscopy (tr-XAS) experiments at the Ni $L_{2,3}$ absorption edges which we analyze quantitatively with \textit{ab initio} time-dependent density functional theory (TDDFT) including local electron correlations.

\begin{figure}
\includegraphics{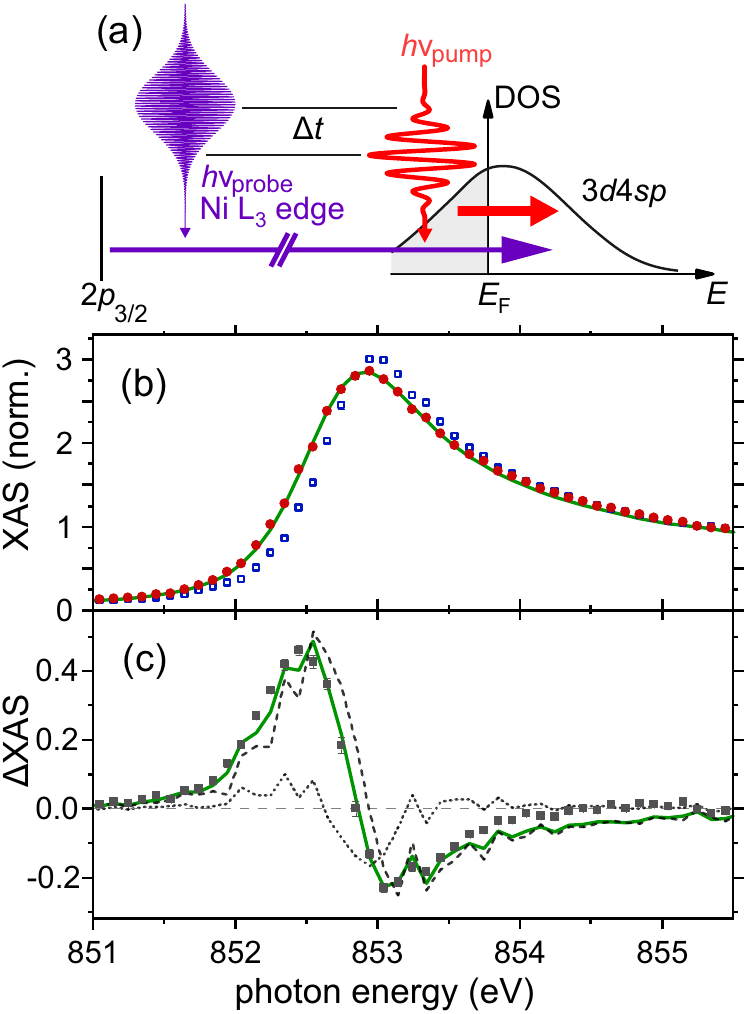}
\caption{(a) Near-infrared pump, soft x-ray absorption probe experiment at the Ni $L_3$ absorption $2p_{3/2}\rightarrow 3d4s$ analyzing the transiently modified electronic density of states above $E_{\mathrm F}$ at time delay $\Delta t$. (b) Ground state (\color{blue}$\square$\color{black}) and pumped (\color{red}$\bullet$\color{black}) absorption spectrum at $\Delta t = 0.4$~ps. The pump-induced changes $\Delta$XAS are modeled based on the static absorption spectrum (green line) which allows to distinguish the contributions of an energy shift and broadening. (c) Pump-induced change ($\blacksquare$) including the modelling result in (b) (green line). The dashed (dotted) line indicates fits with only an energy shift (broadening), which are insufficient to describe the data.}
\label{fig1}
\end{figure}

Fig.~\ref{fig1}(a) sketches the pump-probe experiment which measures the temporal correlation of ultrashort X-ray probe pulses tuned to the Ni $L_{2,3}$ edges with near-infrared pump pulses of photon energy $h\nu=1.5$~eV, 35~fs duration, and 12~mJ/cm$^2$ incident fluence as a function of time delay $\Delta t$. The core level resonance involves a transition from $2p_{3/2}$ ($2p_{1/2}$) to $3d4sp$ final states at the $L_3$ ($L_2$) edge. Thereby, we analyze the effect of the optical excitation on the unoccupied $3d4sp$ electronic density of states (DOS) through the time-dependent absorption changes.

The experiments were performed at the Spectroscopy and Coherent Scattering Instrument (SCS) of European XFEL (EuXFEL) \cite{Tschentscher2017,decking2020}. The spectra were measured using linearly polarized monochromatic X-ray pulses with $\Delta E/E = 5\cdot10^{-4}$ \cite{gerasimova2022} tuned between 840 and 880~eV to cover the $L_{2}$ and $L_{3}$ absorption edges \cite{note_data}. The employed X-ray delivery time pattern consisted of fifty 50~fs X-ray pulses in one train with a train repetition rate of 10~Hz and an intra-train repetition rate of 70~kHz. The pump pulses \cite{Tschentscher2017,Pergament2016} were synchronized with every second X-ray pulse. A transmission zone plate \cite{Doring2020} splits the incoming X-rays into three focused, spatially distinct beams of equal intensity in diffraction orders -1, 0, 1. This setup allowed the simultaneous detection of the pumped, the unpumped, and the reference signal for the identical X-ray pulse, which is essential at a SASE free electron laser due to fluctuations in the intensities of subsequent pulses and was made possible by a MiniSDD-based DSSC detector, a 1Mpixel camera with a peak frame rate up to 4.5~MHz \cite{porro2021}. The time resolution of the experiment was 80~fs full width at half maximum (FWHM). Polycrystalline fcc Ni film samples were evaporated on a $5\times5$ array of 200~nm thick Si$_{3}$N$_{4}$ windows, with the middle window being left uncovered for the reference measurement. They consist of a 20$\pm$0.7~nm thick fcc Ni layer, capped by 2~nm MgO. A 100~nm thick Cu layer was deposited on the backside of the window to mitigate heating effects. The samples are polycrystalline and ferromagnetically ordered \cite{Rothenbach2020}. The XAS spectra were recorded at room temperature in transmission geometry and evaluated using a dedicated toolbox \cite{Tschentscher2017,Fangohr:ICALEPCS2017-TUCPA01,castoldi2019,fangohr:icalepcs2019-tucpr02,LoicToolBox}. The pump-induced change is calculated as the negative logarithm of the pumped signal divided by the unpumped in combination with flat-field and non-linearity corrections \cite{LoicToolBox,heat}. The spectra are corrected with a linear background and are normalized according to the edge jump \cite{supp}, which allows a quantitative comparison of experiment and theory.

Fig.~\ref{fig1}(b) depicts the ground state (unpumped) and pumped fcc Ni $L_3$ edge spectra at $\Delta t=0.4$~ps, Fig.~\ref{fig1}(c) the pump-induced change of the absorption spectrum. A positive change is observed at the rising edge around 852.0-852.8~eV, followed by a smaller negative change at 853-854~eV. To model these changes, we need to account for both a spectral redshift and a broadening by modifying the unpumped spectrum, respectively, with a rigid redshift and a broadening via convolution \cite{supp}. We quantify the redshift to $104\pm 25$~meV and the broadening to $139\pm 10$~meV, which we assign in the following to changes in the electronic structure and electron redistribution, respectively. The latter can be intuitively understood from the excitation of holes (electrons) below (above) $E_{\mathrm F}$ by the pump pulse, see Fig.~\ref{fig1}(a).

\begin{figure}
\includegraphics{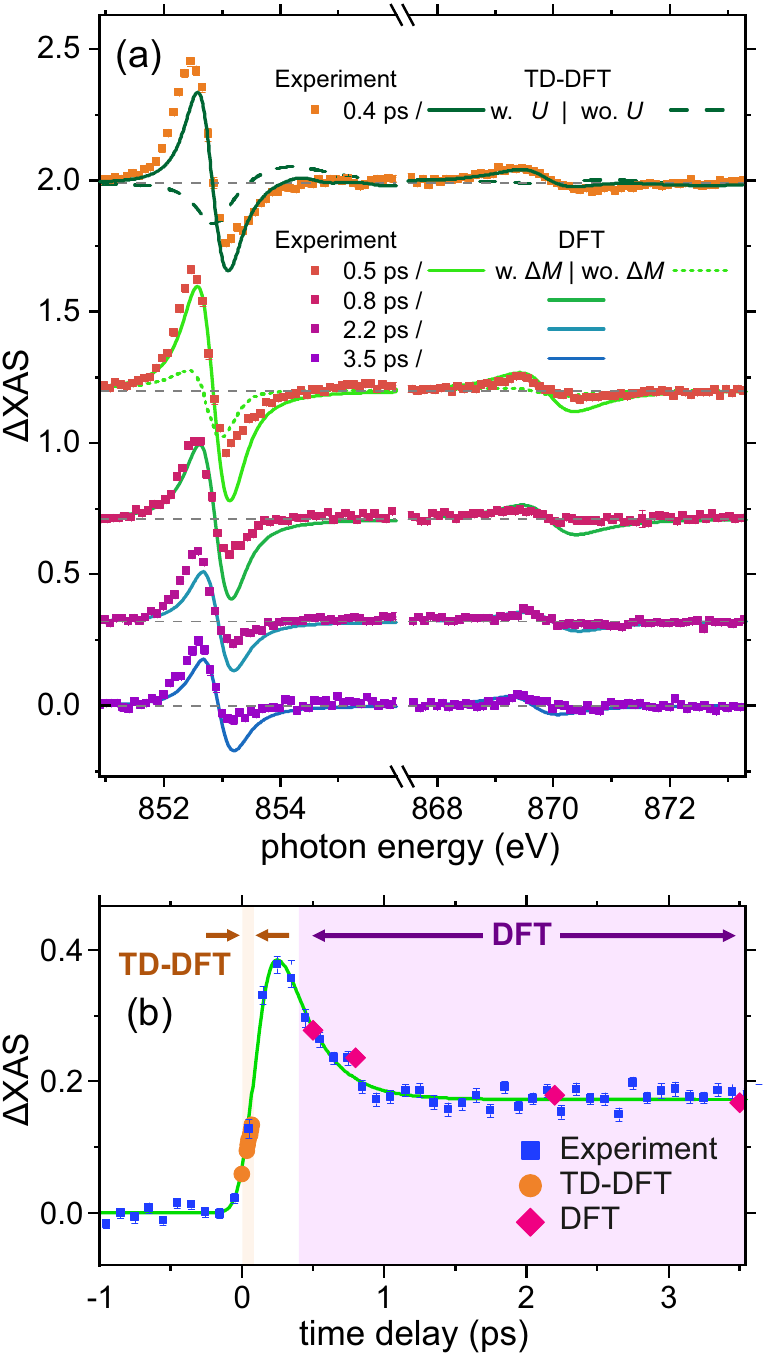}
\caption{a) Pump-induced changes $\Delta$XAS at the indicated time delays from experiment (markers) and TDDFT respectively DFT calculations (solid lines). For comparison, TDDFT calculations without local correlations (dashed line) and DFT calculations without (dotted line) a reduced magnetization (see text for details), are shown. (b) Time-dependent $\Delta$XAS at $h\nu=852.72$~eV with a fit (green line) and the corresponding values from TDDFT (convoluted with a Gaussian of 80~fs FWHM) and DFT, as indicated. }
\label{fig2}
\end{figure}

Fig.~\ref{fig2}(a) details the spectral dependence of the pump-induced change for 0.4~ps$\leq\Delta t\leq$3.5~ps at both absorption edges. We find that the positive change at lower photon  energy recedes to about half within 3.5~ps while keeping its overall shape. The $L_2$ edge generally exhibits a smaller and energetically broader change, which we explain by the larger lifetime broadening at the $L_2$ compared to the $L_3$ edge \cite{supp}.

We now look at the time dependence in more detail by scanning $\Delta t$. In Fig.~\ref{fig2}(b), we show the evolution of the absorption change at a constant $h\nu=852.72$~eV. Thereby we confirm the absorption change of up to 0.4 at fixed $h\nu$ reported in Fig.~\ref{fig2}(a). These time-dependent data highlight that the large change occurs within 200~fs after pumping, while the excess energy resides mostly in the electronic system. The experimental data were fitted with exponential rise and decay times $\tau_{1,2}$, respectively, convoluted with a Gaussian of 80~fs FWHM to account for the time resolution \cite{supp}. We find $\tau_1=131\pm10$~fs and $\tau_2=231\pm7$~fs, which we assign to electron thermalization and electron-phonon coupling in good agreement with previous work \cite{stamm_femtosecond_2007,stamm_2010}.

For a theoretical analysis of the optically induced non-equilibrium state we employ TDDFT, which extends the ground state density functional theory (DFT) to the time domain through the exact one to one correspondence between the time-dependent external potential and the density \cite{runge_density-functional_1984}. The time-dependent Hamiltonian of an  interacting system is mapped onto an equivalent non-interacting one known as the time-dependent Kohn-Sham (TDKS) Hamiltonian with an effective Kohn-Sham (KS) external potential that produces the same density of the interacting system \cite{supp}. This allows to simulate the dynamics of matter subject to a time-dependent perturbation, e.g., the effect of an optical pulse on the electronic structure \cite{castro_optical_2004,bertsch_real-space_2000}. A general approach for calculating time-dependent XAS using a mixed scheme between the linear response of TDDFT and the time evolution of the TDKS is outlined in Ref. \cite{dewhurst_element_2020} and the static response function $\chi_0$ of the KS quasiparticles is given in the random phase approximation by
\begin{equation}
\chi_0(\omega)=\lim_{\eta \to 0}\sum_{ijk}(f_{ik}-f_{jk})\frac{\phi_{ik}^\ast (\mathbf{r})\phi_{jk}^\ast(\mathbf{r}^{'})\phi_{ik}(\mathbf{r}^{'}) \phi_{jk}(\mathbf{r})}{\omega-(\epsilon_i-\epsilon_j)+i\eta}.
\label{eq:non-interacting-chi}
\end{equation}
Here $f_{ik}$ is the occupation of the KS state, $\phi$ is the single particle KS state, $i,j$ are band indices, $k$ is the electron's crystal momentum, $\eta$ is proportional to lifetime broadening, and $\epsilon_i$ is the KS energy \cite{petersilka_excitation_1996}. This approach has previously been used to provide a qualitative description of time-resolved x-ray magnetic circular dichroism spectra using only the transient KS populations \cite{yao_distinct_2020,dewhurst_element_2020}. Here, we use the full transient quantities in Eq.~\ref{eq:non-interacting-chi}, namely occupations, energies, and KS orbitals projected on the ground state, following \cite{elhanoty_2022}. We introduce the electronic correlations to our system from the Hubbard model and consider a Hamiltonian of the form 
\begin{equation}
  \hat{H}=\hat{H}_0(t)+U\sum_i n_{i\uparrow}(t)n_{i\downarrow}(t),
    \label{eq:Hubbard model}
\end{equation}
where $\hat{H}_0$ is the quasiparticle Hamiltonian assumed to be equivalent to the KS Hamiltonian in the Local Spin Density Approximation (LSDA), $U$ is the onsite Hubbard correlation, and $n_{\uparrow,\downarrow}$ are the number operators of spin up and spin down electrons, respectively. The response function corresponding to $\hat{H}$ in mean field (Hartree Fock) solution is
\begin{equation}
\chi_0^H(\omega)=\lim_{\eta \to 0}\sum_{ijk}(f_{ik}-f_{jk})\frac{\phi_{ik}^\ast (\mathbf{r})\phi_{jk}^\ast(\mathbf{r}^{'})\phi_{ik}(\mathbf{r}^{'}) \phi_{jk}(\mathbf{r})}{\omega-(\epsilon_i-\epsilon_j+U\cdot m)+i\eta},
\label{eq:non-int-chi-with-correlation}
\end{equation}
where $m=\langle n_\uparrow \rangle -\langle n_\downarrow \rangle$ is the magnetization. Eq. \ref{eq:non-int-chi-with-correlation} implies that the optical excitation of an initial ground state ($\epsilon_i-\epsilon_j$), which is accompanied by spin flips,  experiences a shift of $U\cdot m$ \cite{elhanoty_2022}.

\begin{figure}
\includegraphics[width=\linewidth]{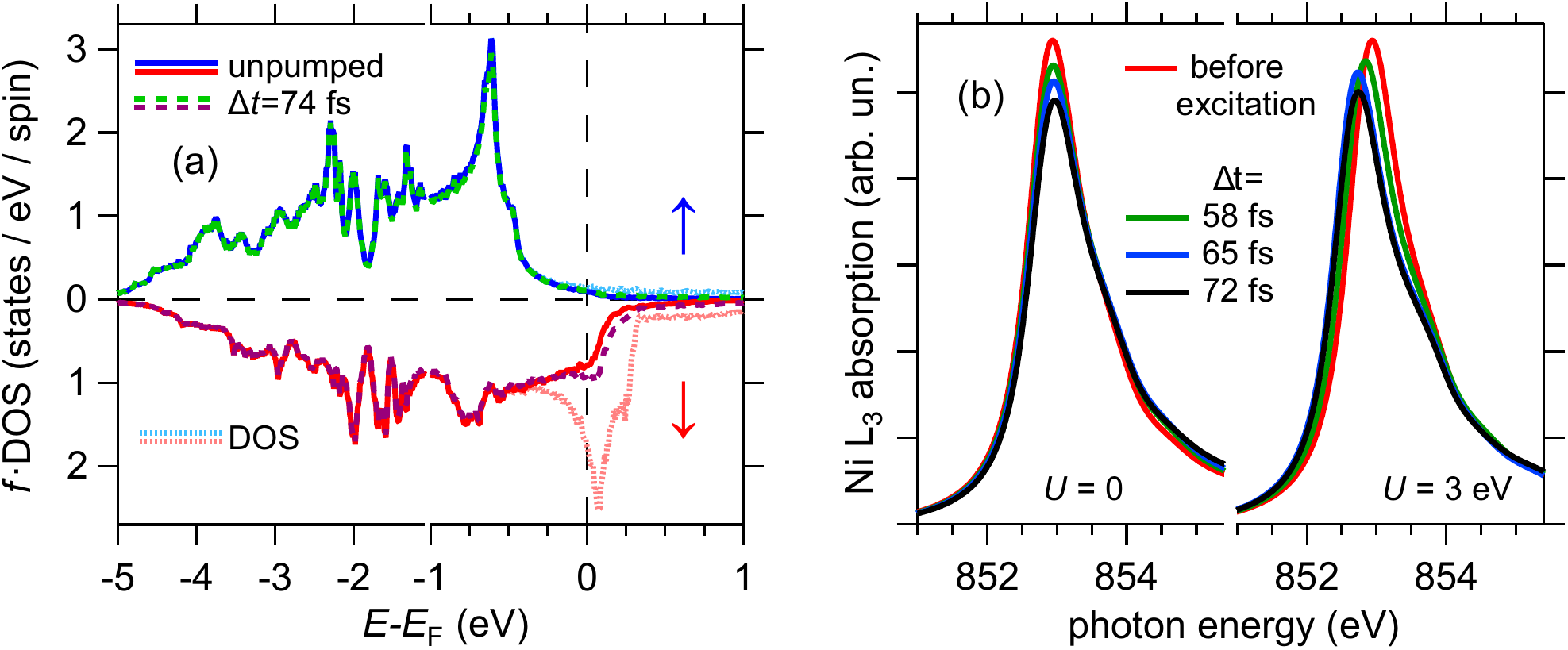}
\caption{(a) Populated exchange-split density of states for KS states in fcc Ni: $f\cdot\mathrm{DOS}$ calculated by TDDFT for majority ($\uparrow$) and minority ($\downarrow$) states before optical excitation (solid lines) and at $\Delta t=74$~fs (dashed lines). The static DOS without population is shown for comparison (dotted lines). (b) Absorption spectrum of the Ni $L_3$ edge after optical excitation calculated by TDDFT using the transient $f\cdot \mathrm{DOS}$ from panel (a) at left and including $U=3$~eV following eqs.~\ref{eq:Hubbard model},\ref{eq:non-int-chi-with-correlation} in addition to $f\cdot \mathrm{DOS}$ at right.}
\label{fig3}
\end{figure}

Before electron thermalization through electron-electron scattering on few 100~fs timescales \cite{rhie_femtosecond_2003, chang_electron_2021}, the transient change in the populated DOS is considered as the product of a time- and energy-dependent non-equilibrium distribution function $f(E,t)$ and an equally time- and energy-dependent DOS$(E,t)$.  Fig.~\ref{fig3}(a) shows the calculated $f\cdot\mathrm{DOS}$ for the experimental conditions before the optical excitation (unpumped) and at $\Delta t=74$~fs after pumping. Upon excitation we find an increase in $3d_{\downarrow}$ orbitals within an interval of $\pm0.2$~eV around $E_\mathrm{F}$ and a weaker decrease in $3d_{\uparrow}$ states in the vicinity of the peak near  $E-E_\mathrm{F}=-0.7$~eV. This behavior is explained by spin-orbit coupling mediated spin currents in the optically excited electron system that induce spin-flip transitions from the majority to the minority channel and lead to a reduced value of $m$ \cite{krieger_2015,toews_pastor_2015}. The transient populations and energies are used as input for calculations of absorption spectra at the Ni $L_{2,3}$ edges following Eq.~\ref{eq:non-interacting-chi} and the results are depicted in Fig.~\ref{fig3}(b), left. The excited electron distribution leads to a reduction of the absorption peak's height, which is observed in the experiment as well, see Fig.~\ref{fig1}(b).  Therefore, the calculated absorption spectra using the time evolution of the TDKS equation within the adiabatic LSDA, presented in Fig.~\ref{fig3}(b) on the left, are well in line with the calculated $f\cdot\mathrm{DOS}$ in Fig.~\ref{fig3}(a), but they lack the spectral shift observed in the experiment, see Fig.~\ref{fig1}. While this approximation successfully captures the changes in the occupations $f_{ik}-f_{jk}$ it is deficient in reproducing the changes in the excitation energies, which signals the influence of electron correlations. To account for these we adopt Eq.~\ref{eq:Hubbard model}, using $U=3$~eV for the on-site correlation \cite{bandyopadhyay_calculation_1989}. The correlations modify the excitation energies of the non-interacting KS response function of Eq.~\ref{eq:non-interacting-chi} by the $U\cdot m$ term, see Eq.~\ref{eq:non-int-chi-with-correlation}. Calculations of the absorption spectrum including $U$ are depicted in Fig.~\ref{fig3}(b) at right and show the transient redshift. Since the reduced $m$ is already obtained in the calculations without $U$ the ultrafast spectral redshift is assigned to the cooperation of $U$ and $m$, leading to changes in the DOS. We note that earlier work using DFT \cite{carva_2009} obtained a spectral redshift without considering either $U$ or the transient spin currents, which determine the ultrafast demagnetization. Such an approach relying on an elevated electronic temperature can only capture the spectral changes partially, and we stress that modifications to the electronic structure as just introduced are essential to describe the experimental data, as will be shown in the following. 

In the top part of Fig.~\ref{fig2}(a) we compare the $\Delta$XAS spectrum calculated by TDDFT with the one measured at the smallest available $\Delta t = 0.4$~ps quantitatively and obtain a very good agreement between experiment and theory. For comparison, we calculated the spectral changes for $U=0$, which is found to be qualitatively different, see Fig.~\ref{fig2}(a), and can indeed not describe the experimental observation.

To describe fcc Ni at $\Delta t > 100$~fs, which represents times too long for TDDFT to be carried out accurately, we approximate the excited state on these timescales by an elevated electron temperature $T_{\mathrm{e}}$ and a reduced $m$ \cite{koopmans_2010} in quasi-static constrained DFT calculations. In the bottom part of Fig.~\ref{fig2}(a) $\Delta$XAS is compared with these DFT results for $\Delta t \geq 0.5$~ps. We find very good agreement with experiment for $T_{\mathrm{e}}$ relaxing from 570~K to 340~K, combined with reduced magnetic moments per atom (compared to the static $\mu=0.61~\mu_\mathrm{B}$ at $T=0$K) of $\mu=0.47~\mu_\mathrm{B}$ to $0.56~\mu_\mathrm{B}$, for 0.5~ps $<\Delta t < 3.5$~ps. Taking only an increased $T_{\mathrm{e}}$ into account and keeping $m$ constant is insufficient to obtain the observed $\Delta$XAS, which highlights the sensitivity of this technique to the changed $m$ even using linearly polarized x-rays.

In Fig.~\ref{fig2}(b) the calculated changes from TDDFT and DFT are shown on top of the time dependent measurement at fixed $h\nu$. TDDFT covers $\Delta t<100$~fs of the initial absorption increase  \cite{note_TDDFT}, reinforcing that the initial non-equilibrium state involves correlation-induced modifications of the electronic structure. After electron thermalization at $\Delta t>400$~fs and for the subsequent cooling of $T_{\mathrm{e}}$ and simultaneous relaxation of the optically induced demagnetization, agreement with our DFT calculations is found. While the two theory data sets complement each other in covering the time interval studied in the experiment, both theory results agree well with the experimental results. On this basis, we assign the experimentally observed, transient spectral broadening to electronic redistribution described by $f(E,t)$. There is reasonable agreement between the experimentally observed spectral broadening of $130\pm 10$~meV and $T_\mathrm{{e}}=570$~K at 0.5~ps, i.e. $\Delta T_\mathrm{{e}}=270$~K above room temperature, considering that $4\cdot k_{\mathrm{B}}\Delta T_\mathrm{{e}}=93$~meV. Deviations between experiment and theory in the negative change of $\Delta$XAS, which get more pronounced with longer $\Delta t$ (compare Fig.~\ref{fig2}(a)), are potentially due to effects not covered in theory, e.g. (non-thermal) phonon transport into the substrate \cite{Rothenbach2019,Rothenbach2021}. 

In conclusion, we present experimental tr-XAS for fcc Ni in the non-equilibrium regime after fs laser excitation, with unprecedented resolution in the time and energy domain, in combination with \textit{ab initio} theory, which allows to identify the optically induced electron repopulation and reduction of the magnetization. Moreover, our combined time and energy resolution explains the transient redshift of the absorption spectrum at the $L_{2,3}$-edges as a signature of electron correlations, as signaled by the Hubbard $U$ and its influence on the electronic response function. The success of our theory on a mean field, Hartree-Fock level put forth here in reproducing experimental observations is encouraging for the analysis of similar systems (intermetallic compounds and alloys) in out-of-equilibrium conditions. Hence this study has potential in offering a more general understanding of the influence of local correlations on the scattering rate, e.g. in Fe and Co \cite{sanchez2012effects}, and spectral weight transfer problems in strongly correlated materials with emergent phases in general. Such dynamics has indeed been discussed recently in work by ultrafast XAS regarding the light induced renormalization of the onsite Coulomb repulsion in a cuprate \cite{baykusheva_2022}. It is also noteworthy that the theoretical analysis presented here does not rely on a renormalized screening of the Hubbard $U$, and is in this sense consistent with recent work on NiO \cite{graanas_PRR22}. Our approach of combining state of the art time and energy resolution in soft x-ray absorption spectroscopy with \emph{ab initio} theory paves the way for full access to the non-equilibrium electronic structure and many-body effects of the broad class of solid materials that exhibit local correlations and magnetic order. \\

\begin{acknowledgments}
We acknowledge European XFEL in Schenefeld, Germany, for provision of X-ray free-electron laser beamtime at the SCS instrument and would like to thank the staff for their assistance. Funding by the Deutsche Forschungsgemeinschaft (DFG, German Research Foundation) - Project-ID 278162697 - SFB 1242 is gratefully acknowledged. The computations were enabled by resources provided by the Swedish National Infrastructure for Computing (SNIC) at NSC and Uppmax partially funded by the Swedish Research Council through grant agreement no. 2018-05973. OG acknowledges financial support from the Strategic Research Council (SSF) grant ICA16-0037 and the Swedish Research Council (VR) grant 2019-03901. This work was also supported by the European Research Council via Synergy Grant 854843 - FASTCORR. O.E. acknowledges support also from eSSENCE, the Knut and Alice Wallenberg foundation, The Swedish Research Council and the Foundation for Strategic Research. C.B. was supported by the Region Grand Est grant 19P07304 - FEMTOSPIN. P.S.M., R.Y.E. and M.B. acknowledge funding from the Helmholtz Association via grant VH-NG-1105. 
\end{acknowledgments}

T. L. performed the experiments and analyzed the data. M. E. developed the extension of TDDFT and did the calculations. Both contributed equally to this work.

\bibliographystyle{apsrev4-2}
%\bibliography{refs_lojewski}

\begin{thebibliography}{48}
\bibitem [{hea()}]{heat}%
  \BibitemOpen
  \href@noop {} {\bibinfo {title} {To avoid effects due to heat build-up in the
  sample within one train, only the first six pulses are evaluated in the
  analysis.}}\BibitemShut {Stop}%
\end{thebibliography}


\end{document}